# Energy Efficient Learning with Low Resolution Stochastic Domain Wall Synapse Based Deep Neural Networks

**Walid A. Misba[1], Mark Lozano[1], Damien Querlioz[2], Senior Member, IEEE, and Jayasimha Atulasimha[1,3], Senior Member, IEEE**

[1]Mechanical and Nuclear Engineering Department, Virginia Commonwealth University, Richmond, VA 23284 USA
[2] Université Paris-Saclay, CNRS, Centre de Nanosciences et de Nanotechnologies, Palaiseau, France
[3]Electrical and Computer Engineering Department, Virginia Commonwealth University, Richmond, VA 23284 USA

Corresponding author: (e-mail:). misbawa@vcu.edu, jatulasimha@vcu.edu

This work was supported in part by the National Science Foundation (NSF) under Grant ECCS 1954589 and Grant CCF 1815033 and in part by Virginia Commonwealth Cyber Initiative (CCI) CCI Cybersecurity Research Collaboration Grant.

**ABSTRACT** We demonstrate that extremely low resolution quantized (nominally 5-state) synapses with large stochastic variations in Domain Wall (DW) position can be both energy efficient and achieve reasonably high testing accuracies compared to Deep Neural Networks (DNNs) of similar sizes using floating precision synaptic weights. Specifically, voltage controlled DW devices demonstrate stochastic behavior as modeled rigorously with micromagnetic simulations and can only encode limited states; however, they can be extremely energy efficient during both training and inference. We show that by implementing suitable modifications to the learning algorithms, we can address the stochastic behavior as well as mitigate the effect of their low-resolution to achieve high testing accuracies. In this study, we propose both in-situ and ex-situ training algorithms, based on modification of the algorithm proposed by Hubara et al. [1] which works well with quantization of synaptic weights. We train several 5-layer DNNs on MNIST dataset using 2-, 3- and 5-state DW device as synapse. For in-situ training, a separate high precision memory unit is adopted to preserve and accumulate the weight gradients, which are then quantized to program the low precision DW devices. Moreover, a sizeable noise tolerance margin is used during the training to address the intrinsic programming noise. For ex-situ training, a precursor DNN is first trained based on the characterized DW device model and a noise tolerance margin, which is similar to the in-situ training. The highest inference accuracies we obtain after the in-situ and ex-situ training are ~ 96.67% and ~96.63% which is very close to the baseline accuracy of ~97.1% obtained from a similar topology DNN having floating precision weights with no stochasticity. For ex-situ inference, the energy dissipation to program the DW devices is calculated to be 2.18 pJ per inference. Remarkably, for in-situ inference the energy dissipation to program the devices is only 13 pJ per inference given that the training is performed over the entire MNIST dataset for 10 epochs. Large inter-state interval provided by the quantized weights and noise tolerance margin allowed during the in-situ training enables it to perform the training with significantly lower number of programming attempts. Our approach is specifically attractive for low power intelligent edge devices where the ex-situ learning can be utilized for energy efficient non-adaptive tasks and the in-situ learning can provide the opportunity to adapt and learn in a dynamically evolving environment.

## I. INTRODUCTION

Deep neural networks (DNNs) have proven to be successful in image recognition and other big data driven classification tasks. However, implementing a DNN with traditional von-Neumann computing is time consuming [2] as it requires shuttling a large number of synaptic weight data stored in the memory to the processing unit to perform matrix-vector multiplication during the forward propagation and backward propagation stages. Moreover, shuttling data between the computational unit and memory unit is energy intensive [3], which hinders the implementation of such DNNs in edge devices where energy is at a premium.

In-memory computing has been widely explored to reduce the physical separation between computation and memory unit. In-memory computing is a non-von-Neumann computing paradigm where the computational memory units are arranged in a way that certain computational tasks take place in the memory itself [4-5]. Matrix vector multiplication, the most computationally intensive part of a DNN [6], has been demonstrated with in-memory computing [7-8]. When the computational memory units are connected in a crossbar and programmed to provide conductances equivalent to the DNN weights [9-10], the matrix-vector multiplication operation can be implemented in single time step [4,6] and with minimal data movement. Computational memory such as phase change random access memory (PCM) [11,12], resistive random-access memory (RRAM) [13,14], arranged in a crossbar array have been shown to classify handwritten digits [9,15] and recognize human faces [16]. However, these analog memory devices have stochastic and non-linear responses and provide limited resolution for synaptic weights. To achieve higher classification accuracy, these issues should be addressed with appropriate training algorithms.

Recently spintronic memory devices are being widely explored for in-memory DNN implementation because of their non-volatility, high endurance, high speed of access, high scalability and compatibility with CMOS technology [2,17,18-22]. Among these spintronic devices, DW based computation memory [17,18] is promising and these devices can be programmed with a low energy budget [23]. However, similar to other analog devices, DW devices have limitations such as their stochastic behavior [24-27] and low resolution due to the relatively small on/off ratio of magnetic tunnel junctions (MTJs) which are 7:1 at best at room temperature [28].

With recent advances in computing, researches have shown fast and energy efficient implementation of DNNs with low resolution synaptic weight [1,29-32], where the weights value can be only binary (1-bit or 2-state) [29]. However, for the weight update, gradients are calculated in full precision to achieve high accuracy [1]. This idea of keeping full precision gradient information for training a network with limited precision synaptic weights can be useful for a DNN that is built from energy efficient DWs or other analog low-resolution devices.

Apart from the low resolution, stochasticity and non-linear response of the analog devices should be addressed during training to achieve higher classification accuracy. To address stochasticity of the analog devices both online (in-situ) and off-line (ex-situ) training of the DNN are proposed. For online training, multiple devices per synapse have been proposed with [33] or without 'periodic carry' [34] to address device variability and noise. In another work, a 3T1C module (consists of 3 CMOS transistors and 1 capacitor) is used in conjunction to the stochastic PCM device to accumulate small linear updates and then periodically transfer them to the non-volatile PCM [35]. However, with online training using the techniques mentioned above during the weight update stage, each of the synaptic weights in the cross-bar array are updated. This has great implications for the endurance of the devices as well as energy consumed in training the device. Recently, mixed precision framework [36-37] has been proposed where large computational load such as the weighted sum operation (matrix-vector multiplication) along with the conductance updates are performed in low precision computational memory unit and the weight updates are accumulated in a high precision unit. Using this framework, a large variety of DNNs have been shown to achieve high classification accuracy with significantly smaller number of weight updates [36].

In contrast to the online training, for offline training the DNN is trained in software and the actual devices are programmed based on the learned weights from software. In this case, hardware nonidealities are characterized first and then included in the training process. To address stochasticity of the devices, Gaussian noise injection for the DNN weights has been proposed [38] and have shown excellent accuracy. Random gaussian noise is also added to the ternary weights (3-state weight) of a DNN [39]. Variation aware off-line training algorithm is reported in [40-41] where the variation in device conductances and device defects are first characterized and then incorporated during the training of the DNN in software. In another case involving a deep convolutional neural network, the optimal weights for convolution layers and fully-connected layers are learned via off-line training before the fully connected layers' weights being updated by online training [42]. In this work, we have shown both online and off-line training strategies to achieve high classification accuracy for DNN implemented with highly stochastic (non-gaussian) and extremely low resolution (nominally 2-state, 3-state and 5-state for synaptic weights) analog DW based computational memory devices. Moreover, we show that our proposed framework for on-line training requires significantly lower number of weight updates.

The rest of the paper is organized as follows. In the methods section, we detail the architecture of the DW device that can work as a synapse in the DNN and discuss the in-situ and ex-situ learning algorithms of such DW device based DNN. For both of the learning algorithms we adapt quantized neural network learning algorithm [1,29] with several modifications including the weight deviation tolerance from target weight to account for the programming noise intrinsic to such stochastic DW devices. For ex-situ training, we also incorporate the statistical distribution of the DW device conductance during the training, which helps to achieve higher test accuracy. This is followed by the results and discussion section, and then a conclusion.

## II. Methodology

### A. DW BASED NANO-SYNAPSE AND MICROMAGNETIC MODELING FOR DEVICE STOCHASTICITY

We model our synapse on a magnetic domain wall (DW) based nanodevice, which is non-volatile in nature. Once the memory state (here the synaptic weight) is written, the information is retained for a long time. For the nano-synapse device, we simulated a thin ferromagnetic racetrack having a dimension of 600 nm × 60 nm × 1 nm with a DW initialized and stabilized in a notch at one end. In addition, we assume several engineered notches at regular intervals along the racetrack. The racetrack dimension and notch intervals are shown in Fig. 1a. Moreover, we considered edge irregularities (rms roughness of ~ 2nm) in the racetrack to mimic the effect of lithographic imperfections by randomly removing or adding some finite difference cells from the edges [43-44]. We assume the racetrack is on top of a heavy metal layer that is patterned on top of piezoelectric layer (see Fig. 1b). An insulator (MgO layer) and a reference ferromagnetic layer (one could also add a synthetic antiferromagnet (SAF) layer to cancel dipole coupling from this fixed layer) are stacked on top of the racetrack, these two layers combined with the racetrack ferromagnetic layer (free layer) forms a magnetic tunnel junction (MTJ), which facilitates the readout of the device. With this configuration, a combination of fixed amplitude and fixed time current pulse "or clocking signal" injected in the heavy metal layer and a varying amplitude "control" voltage pulse applied across the piezoelectric translates the domain wall into different distances along the racetrack. Different positions of the DW lead to different conductances of the MTJ (see Fig. 1b) thus forming a voltage programmable non-volatile state.

### B. MAGNETIZATION DYNAMICS

The magnetization dynamics in the presence of heavy metal layer current, which exerts Spin Orbit Torque (SOT) on the ferromagnetic strip is simulated in MUMAX3 [45] using the Landau–Lifshitz–Gilbert-Slonczewski equation:

$$(1 + \psi^2)\frac{d\vec{m}}{dt} = -\gamma\vec{m} \times \vec{H}_{eff}$$
$$- \psi\gamma\left(\vec{m} \times \left(\vec{m} \times \vec{H}_{eff}\right)\right)$$
$$- \beta\gamma(\varepsilon - \psi\varepsilon')(\vec{m} \times (\vec{m}_P \times \vec{m}))$$
$$+ \beta\gamma(\varepsilon' - \psi\varepsilon)(\vec{m} \times \vec{m}_P) \qquad (1)$$

$$\beta = \frac{\hbar J\theta}{\mu_0 e d M_s},$$
$$\varepsilon = \frac{P\Lambda^2}{(\Lambda^2 + 1) + (\Lambda^2 - 1)(\vec{m} \cdot \vec{m}_p)} \qquad (2)$$

We consider secondary spin torque parameter to be $\varepsilon' = \psi\varepsilon$ and neglect the field like torque. Here, spin polarization $\vec{m}_P = \hat{J} \times \vec{z}$ , where $\hat{J}$ is the unit vector defining the direction of current flow and $\vec{z}$ is the direction of inversion asymmetry. Also, $J$ is the value of current flowing through the heavy metal layer, $\theta$ is the spin Hall angle which is 0.1 for heavy metal Pt, $\psi$ is the damping constant, $\gamma$ is the gyromagnetic ratio, $\vec{m}$ is the unit magnetization vector, $M_s$ is the saturation magnetization, $\hbar$ is the reduced Planck constant, $\mu_0$ is the permeability of free space, $e$ is the electron charge and $d$ is the thickness of the nanowire. To equate the Slonczewski torque with Spin orbit torque we assume spin polarization to be $P = 1$ and Slonczewski parameter to be $\Lambda = 1$. Here the effective field, $\vec{H}_{eff}$ accounts for the contributions from demagnetization, perpendicular magnetic anisotropy (PMA), exchange interaction from Heisenberg and Dzyaloshinskii–Moriya interaction (DMI), stress induced anisotropy and thermal noise. $\vec{H}_{eff}$ can be expressed as follows:

$$\vec{H}_{eff} = \vec{H}_{anis} + \vec{H}_{demag} + \vec{H}_{stress} + \vec{H}_{exch}$$
$$+ \vec{H}_{thermal} \qquad (3)$$

PMA induced effective field can be expressed as, $\vec{H}_{anis}$ :

$$\vec{H}_{anis} = \frac{2K_u}{\mu_0 M_s}(\vec{u}.\vec{m})\vec{u} \qquad (4)$$

where $K_u$ is the first order anisotropy constant and $\vec{u}$ represents the uniaxial anisotropy direction (i.e. perpendicular to plane). Voltage applied across the piezoelectric induces an in-plane stress [27,46], which is then transferred to the racetrack (heavy metal layer is thin ~ 5nm) and modulates the PMA by means of stress induced anisotropy resulting from the magnetoelastic effect. Voltage induced anisotropy modulation is incorporated by varying the PMA coefficient, $K_u$ of the racetrack in our simulation.

The effective field due to the interfacial DMI is expressed as follows [45]:

$$\vec{H}_{DMI} = \frac{2D}{\mu_0 M_s}\left(\frac{\partial m_z}{\partial x}, \frac{\partial m_z}{\partial y}, -\frac{\partial m_x}{\partial x} - \frac{\partial m_y}{\partial y}\right) \qquad (5)$$

Here, D is the DMI constant and $m_x$, $m_y$ and $m_z$ are the x, y and z component of unit magnetization vector $\vec{m}$ respectively. Thermal noise induces a random effective field $\vec{H}_{thermal}$ [47]:

$$\vec{H}_{thermal} = \vec{\eta}\sqrt{\frac{2\psi kT}{\mu_0 M_s \gamma \Omega \Delta}} \qquad (6)$$

Here, $\vec{\eta}$ is a random variable with Gaussian distribution with mean zero and unit variance and independent (uncorrelated) in each of the 3 cartesian coordinates generated at each time step, $k$ is Boltzmann constant, $\Omega$ is the cell volume, $\Delta$ is the time step size. The simulation parameters are listed in Table I.

## TABLE I
### SIMULATION PARAMETER

| Parameters | Values |
|---|---|
| DMI constant (D) | $0.0006\ Jm^{-2}$ |
| Gilbert damping ($\psi$) | 0.03 |
| Saturation magnetization ($M_s$) | $10^6\ Am^{-1}$ |
| Exchange constant ($A_{ex}$) | $2 \times 10^{-11} Jm^{-1}$ |
| Saturation magnetostriction ($\lambda_s$) | $250\ ppm$ |
| Perpendicular Magnetic Anisotropy ($K_u$) | $7.5 \times 10^5\ Jm^{-3}$ |

### C. MAPPING DOMAIN WALL POSITION TO CONDUCTIVITY

The distribution of equilibrium DW positions for five different programming voltages, represented by different PMA coefficient, $K_u$, in addition to fixed amplitude and fixed time SOT current pulse ($35 \times 10^{10}\ A/m^2$ applied for 1 ns) in the presence of room temperature thermal noise are shown in Fig. 1c. The mean equilibrium DW positions are different for different $K_u$, which implies that different programming voltages can be chosen for different synaptic states. For example, one can select five, three, or two different programming voltages to implement a 5-state, 3-state or 2-state synapse. Equilibrium DW positions can be linearly mapped to a conductance value by means of the following equations:

$$G^{synapse} = \frac{G_{max} + G_{min}}{2} + \frac{G_{max} - G_{min}}{2} < m_z > \qquad (7)$$

where, $< m_z >$ is the average magnetization moment of ferromagnetic racetrack along z-direction. The distribution of $< m_z >$ is shown in Fig. 1d, which can be derived directly from DW position. The maximum conductance $G_{max}$ of the DW nano-synapse device occurs when the average magnetization of the ferromagnetic racetrack is $< m_z >=1$. In this scenario, the DW moves to one end of the racetrack and the magnetization in the racetrack points to the +z-direction (reference ferromagnetic layer magnetization is assumed to point upward, parallel state). The device conductance is minimum $G_{min}$ when the average magnetization is $< m_z >= -1$, the DW moves all the way to the other end and the magnetization in the racetrack points downward (reference layer magnetization points upward, anti-parallel state).

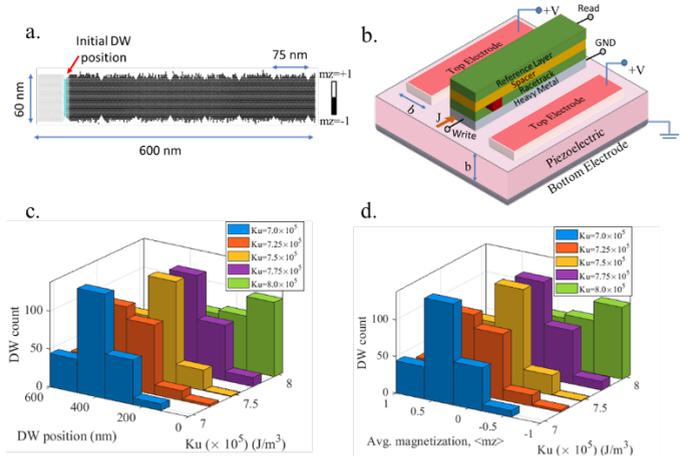

FIGURE 1. a. Micromagnetic configuration of a ~2 nm rms rough edge racetrack with perpendicular magnetic anisotropy (PMA). Engineered notches are placed regularly at 75 nm interval. A DW is initialized at a notch 60 nm from the left of the racetrack. b. DW based nano-synapse device: racetrack ferromagnet/insulator/reference ferromagnet (MTJ) on top of a heavy metal layer on a piezoelectric substrate. A fixed current, J in a heavy metal layer along with different amplitude voltage, V applied across the piezoelectric, which changes the perpendicular anisotropy (PMA or $K_u$ constant) of the racetrack and translates the DW (shown in red rectangle) into different longitudinal positions along the racetrack. c. Distribution of equilibrium DW positions in the racetrack (shown in Fig. 1a) at room temperature T=300K for a fixed SOT current pulse of $J = 35 \times 10^{10}\ A/m^2$ for 1 ns and five different PMA coefficients, $K_u$ (corresponds to five different programming voltages). Different mean positions for different $K_u$ implies that 5-state, 3-state or 2-state stochastic synapses can be implemented by choosing 5,3 or 2 different programming voltages. d. distribution of average perpendicular magnetization, $< m_z >$ (which is equivalent to DNN weights according to Eq. 13) derived directly from DW position.

## III. LEARNING OF FULLY CONNECTED DNN WITH DW NANO-SYNAPSE

### A. CROSSBAR WITH DW DEVICES

We assume a crossbar architecture for the DW devices (Fig. 2b) that implements a fully connected DNN (Fig. 2a). The task of DNN studied here is classification of handwritten digits from MNIST test images [48]. The network is trained with MNIST images each having $28 \times 28$ pixels or a total of 784 pixels with intensity values ranging from 0-255. The pixel intensities of the image, $\{x_1, x_2,..., x_{784}\}$, acts as input of the DNN. The output of the DNN is the classification of the digit for the given input image. For input digit 0 the output should be $\{1,0,0,0....\}$ and for digit 1 the output should be $\{0,1,0,0....\}$ and so on. We have considered 3 hidden layers for the DNN and the numbers of neurons for input layer, hidden layers and output layer are chosen to be 784-392-196-98-10. The reason for the choices is discussed in the results section. In a typical DNN shown in Fig. 2a, at each forward pass, any layer neuron computes the weighted sum of inputs from the previous layer neurons and then passes it through a non-linear activation function to generate its output. If the inputs from layer$-l$ neurons are represented by $x_i^l$, and the weights of the connections between layer$-l$ neuron "$i$" and layer$-(l+1)$ neuron "$j$" is represented by $W_{ij}$, then the output of neuron $j$, is expressed as [49],

$$x_j^{l+1} = f\left(\sum W_{ij} x_i^l\right) \qquad (8)$$

where $f$ represents the activation function of the neuron, for which we used the sigmoid function in our simulation.

In the hardware implementation of DNN with crossbar (see Fig. 2b), the inputs to any layer neurons are transformed into a physical voltage before feeding into the rows of the crossbar. These voltages can be generated by scaling the inputs: $V^{scale}\{x_1^l, x_2^l, \dots, x_i^l, \dots\}$. DW devices situated at each cross-point of the crossbar implements the weights of the connection between two layers of neurons. The conductance of the devices can be scaled linearly to represent the weights $W_{ij}$ of the DNN [17].

$$G_{ij}^{synapse} = \frac{G_{max} + G_{min}}{2} + \frac{(G_{max} - G_{min})W_{ij}}{2\,W_{max}} \qquad (9)$$

Here, $W_{max}$ is the maximum absolute value for the weights of the DNN. DNN weights, $W_{ij}$, can be both positive and negative; however, the DW devices can only provide positive conductance values. To address the issue, one can add a parallel conductance, $G_P = \frac{G_{max}+G_{min}}{2}$, to each of the cross-points in the crossbar and feed this parallel conductance with a voltage that is of opposite polarity to the voltage applied to the DW device [17]. With this parallel conductance the effective conductance, $G_{ij}$, at each cross-point would be,

$$G_{ij} = \frac{(G_{max} - G_{min})W_{ij}}{2\,W_{max}} \qquad (10)$$

The value $G_{ij}$ could be both positive and negative depending on the values of the $W_{ij}$ of the DNN. Once the device conductance, $G_{ij}$, is programmed according to $W_{ij}$, the weighted sum operation, $\sum W_{ij} x_i^l$ (shown in Eq. 8), can be implemented in the crossbar by means of Ohm's and Kirchhoff's laws as follows:

$$I_j = \sum I_{ij} = V^{scale} \frac{(G_{max} - G_{min})}{2\,W_{max}} \sum W_{ij} x_i^l \qquad (11)$$

The additional scaling factor, $V^{scale}\frac{(G_{max}-G_{min})}{2\,W_{max}}$, can be normalized by appropriate design of the peripheral circuitry of the crossbar. In addition, the peripheral circuit computes the activation function and generates the output of the neuron (typically voltage signals), which then acts as inputs for subsequent layer neurons.

When a parallel conductance, $G_P$, is added to the DW device with an input voltage of opposite polarity then, from Eq. 7, the device conductance can be expressed in terms of average magnetization $<m_z>$.

$$G^{synapse} = G_{ij} = \frac{G_{max} - G_{min}}{2} <m_z> \qquad (12)$$

Combining Eq. 10 and 12 and considering the maximum absolute value of DNN weight as $W_{max} = 1$, one can get,

$$W_{ij} = <m_z> \qquad (13)$$

From the above equation it is clear that if we train a DNN shown in Fig. 2a with weights (both positive and negative) that are derived from micro-magnetics (see Fig. 1d) for different programming conditions we are effectively implementing a hardware DNN with DW devices shown in Fig. 2b given that the peripheral circuitry is designed to provide the appropriate scaling.

After each forward pass, the DNN may not be able to correctly predict the input image. Therefore, an error signal is calculated at the output layer, L, which is expressed as $\delta_i^L = y_i^L - d_i^L$, where $y_i^L$ and $d_i^L$ are the predicted and desired outcomes of the output layer's neuron $i$. This error signal is used to train the network's learnable parameters, such as weights, using back propagation of errors and stochastic gradient descent. During the backward pass, any layer's neurons' errors are scaled to voltage values by the peripheral circuitry designed, which are then fed into the cross-bar columns (different from the forward pass where the voltages are fed into the rows of the crossbar) to generate errors for the preceding layer neurons at the rows. In standard backpropagation of a DNN, errors of layer$-$ $l+1$ neurons are back propagated to compute error of the previous layer $-$ $l$ neuron $i$ as follows:

$$\delta_i^l = \sum W_{ij} \delta_j^{l+1} \qquad (14)$$

Weighted sum operation shown above is implemented in crossbar by means of Ohm's and Kirchhoff's laws:

$$I_i = \sum I_{ij} = V^{scale} \frac{(G_{max} - G_{min})}{2\,W_{max}} \sum W_{ij} \delta_j^{l+1} \qquad (15)$$

Similar to the forward pass, in backward pass the peripheral circuit can be designed to normalize the scaling factor. The detailed learning algorithms are discussed in the next section.

## B. BACKPROPAGATION AND LEARNING ALGORITHM

For the training of the DNN, we update the weights by calculating the gradient of a cost function with respect to the weights. We considered mean square error, $C = \frac{1}{2}\sum(y_i^L - d_i^L)^2$ as our cost function where the gradient of the cost function with respect to the output of the output layer neuron $i$ is expressed as , $\delta_i^L = (y_i^L - d_i^L)$ (we also call it error ). Once the output layer's errors are determined, the preceding layer's errors can be calculated using the backpropagation equation, $\delta_i^l = W_{ij}\delta_j^{l+1}$, which is different from the backpropagation equation, $\delta_i^l = W_{ij}\delta_j^{l+1} f'_{l+1}$ reported in ref. [50] where $f'_{l+1}$ is the gradient of the activation function of layer $l+1$ neuron. In other words, we do not back propagate the gradients of activation function as it does not achieve high testing accuracy with quantized weights. Finally, the derivative of the cost function with respect to the weights is calculated, which determines the weight update signal, $\Delta W_{ij}$ for the weights connected between layer $l$ neuron $i$ and layer $l+1$ neuron $j$,

$$\Delta W_{ij} = \eta x_i^l \delta_j^{l+1} f'_{l+1} \qquad (16)$$

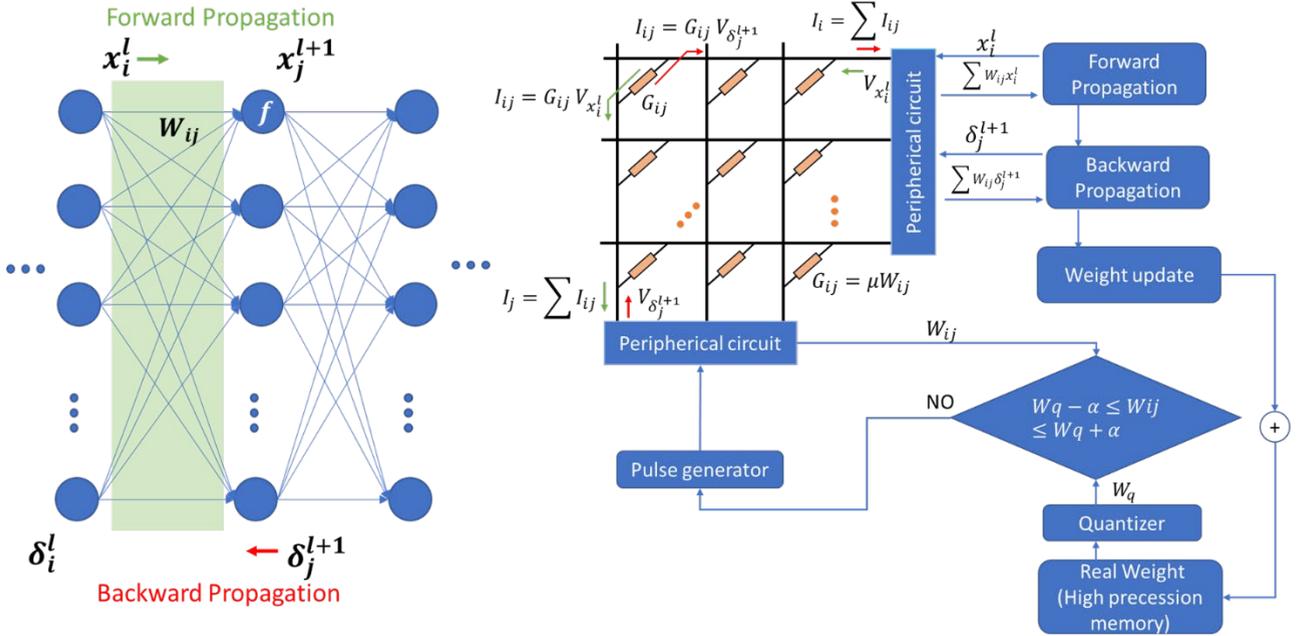

FIGURE 2. a. Architecture of a fully connected deep neural network (DNN). Any neuron $i$ in layer $l$ is connected to neuron $j$ in layer $l+1$ with synaptic weight $W_{ij}$. At forward propagation, inputs to neuron $j$ are summed and passed through an activation function $f$ to generate its output, $x_j^{l+1}$. At backward propagation, errors of layer $l+1$ neurons are back propagated to calculate the error $\delta_i^l$ of neuron $i$ in layer $l$  b. Implementation of the DNN in crossbar with DW devices. The peripheral circuit and the crossbar shown here implements DNN functionalities of only one layer ("$l$") and the next layer ("$l+1$") and the number of rows in the crossbar are determined by the number of neurons in layer, $l$ and the number of columns by the number of neurons in layer, $l+1$. At each cross point of the crossbar there is a DW device with conductance $G_{ij}$, which is equivalent to the DNN weights $W_{ij}$ such that $G_{ij} = \mu W_{ij}$. Inputs and errors of neurons are scaled to voltages before feeding them into the crossbar. The flow of the training algorithm is shown at the right-hand side of the crossbar. For each of the DW devices there is a corresponding high precision weight (real weight) that is stored in a separate memory unit. These high precision weights are updated after a forward and backward pass before passing it through a quantizer (i.e., 2,3 or 5-level quantization, depending on the states of the device). The DW device conductance, $G_{ij}$ (or the corresponding $W_{ij}$) are updated when they fall outside of the prescribed range of the target quantized weight, $W_q$. Figure idea adopted from [36].

Here, $\eta$ denotes the learning rate. For our learning algorithm we propose to store the updated weights in a separate high precision memory unit. That way, the gradients with respect to the weights can be calculated accurately [1]. We note that these high precision weights are different from the actual synaptic weights (or equivalent conductances) provided by the DW device that are quantized and of low precision. However, we use these high precision weights to update the DW device weights (conductances). As we apply stochastic gradient decent for optimization, these high precision weights are updated at each forward pass with an input image.

As DW devices can only provide limited resolution in their synaptic weights we adopt weight quantization in our training algorithm. For that, the high precision weights are quantized at each forward pass during the training. For weight quantization, we use the following sets of functions in the manner of Ref [51]:

$$clip(x, a, b) = \min(\max(x, a), b)$$

$$\Delta = \frac{b-a}{n-1} \qquad (17)$$

$$q = \left[ round\left( \frac{clip(x, a, b) - a}{\Delta} \right) \right] \times \Delta + a$$

Where $q$ is the quantized value of the real valued number $x$, $[a; b]$ is the quantization range and $n$ is the level of quantization. After quantization, a programming pulse is generated to update the DW device weights to the quantized value, a target that is similar to the quantized neural network learning algorithm [1]. Typically, two types of training are possible for a DNN implemented with DW based nano-synapse device: in-situ and ex-situ.

In in-situ training the DNN is trained and tested in hardware. In contrast, in ex-situ training, a precursor DNN is first trained in software and then the learned weights are transferred to the DW devices to provide the equivalent learned weights prior to testing.

### 1) IN-SITU TRAINING

Here, we describe in detail the step-by-step in-situ training algorithm as shown in Fig. 2b. For each DW device in crossbar there is a corresponding high precision weight that is stored in a separate memory unit. Initially, these high precision weights are chosen at random from a gaussian distribution. After each forward and backward pass these weights are updated according to Eq. 16. Then, these weights are clipped and quantized so that they lie between -1 to 1. After that, a programming pulse is sent to the DW device to update its synaptic weight value to the quantized value. For example, in 5-level quantization (5-state for a synaptic device) the quantized weights can be of any value from the set $W_q \in (-1, -0.5, 0, 0.5, 1)$. Five different programming voltages can be applied to the device, which results in $K_u = 8$, 7.75, 7.5, 7.25 and 7.0 ($\times 10^5$) J/m³ to achieve five different quantized weights of -1, -0.5,0,0.5 and -1 respectively as seen from Fig. 1d (DW device weights, $W_{ij} = <m_z>$ according to Eq. 13). Because of the significant spread that exists in the DW device weights (or the $<m_z>$ distribution) due to the stochastic nature of the device we introduced a noise tolerance hyperparameter called alpha $\alpha$ (real valued) during training, as after applying a programming pulse (fixed current + control voltage) the device weights can be of a value other than the desired quantized weight. For instance, if we want to program a DW device to a quantized weight of $W_q$ then we would allow any values for the device weights that satisfy the condition, $W_q - \alpha \leq W_{ij} \leq W_q + \alpha$. Therefore, at each iteration following quantization, we read the states of the DW device (costs read energy but that is typically much lower than write energy) and only if it falls outside the noise tolerance margin, a programming pulse is sent to the device to write the corresponding quantized weight. However, due to the large inter-state interval in quantized learning, a quantized weight does not change at each forward pass (the backpropagated errors update the weights slowly due to low learning rate). Instead, it typically changes only after several passes. Therefore, the noise tolerance condition need not to be satisfied strictly at each iteration. Furthermore, if a DW device weight is programmed outside the tolerance margin, it is not rectified in the current iteration, as it has a chance to satisfy the window in the next several iterations. This relaxation over noise tolerance condition speeds up the training process without losing accuracy. Again, the DW device, which already satisfies the tolerance margin, need not be programmed for next several iterations due to same reason of the quantized weights not being updated frequently. Introduction of noise tolerance hyperparameter, $\alpha$ is critical during the training of this stochastic device based DNN. Without $\alpha$, the DW device needs to be programmed a significantly large number of times to achieve a *particular* quantized weight. On the other hand, a high value of $\alpha$ allows more imprecise weight update or higher variation of the DW device weights from the target values, which will degrade the accuracy. Thus, a proper balance needs to found for selecting the value of $\alpha$ so that it not only ensures high classification accuracy but also low programming energy. Once the DNN is trained, the learned DW device's weights (or conductances) remains the same during testing, as these devices are non-volatile.

We have chosen two representative noise toleration limits for our study that are $\alpha = 0.15$ and $\alpha = 0.25$ . Studies have shown that during training a Gaussian noise of standard deviation, $\sigma$ that is up to 7.5% of the maximum magnitude of DNN weights does not degrade test accuracies significantly when no inference noise is assumed [38]. This motivates us to consider a noise tolerance of $\alpha = 0.15$ that is 15% of the maximum DNN weights (most of the weights in Gaussian distribution lies within $2\sigma \sim 15\%$). However, the DW device we studied here does have inference noise due to the device stochasticity. Furthermore, we choose a maximum noise tolerance of $\alpha = 0.25$ that is 25% of the maximum DNN weights so that the state overlaps between two adjacent states can be restricted for 5-state networks (half of the interstate interval for 5-state is 0.25).

We note that for 3-level quantization (3-state device), DW device can be programmed with control voltages to generate PMA of $K_u = 8$, 7.5, and 7.0 ($\times 10^5$) J/m3 that can achieve quantized weights of -1, 0, and 1 respectively. For 2-level quantization (2-state device), the devices can be programmed to $K_u = 8$ and 7.0 ($\times 10^5$) J/m3 to achieve weights of -1 and 1 respectively. During in-situ training, the device weights are selected randomly from the $<m_z>$ distribution of corresponding $K_u$ to program the DW device to a target quantized value. Although we have computed 250 instances for each of the programming conditions (in Fig. 1c-d) due to the limitation in computational resources, we note that there are dominant pinning sites in the racetrack because of the notches. As a result, the DWs tend to be stuck in or close to those pinning sites in most cases rather than the pinning sites offered by the rough edges of the racetrack (see supplementary Fig. S2). Hence, generating more instances will likely follow the probability distribution, which already exists in the current distribution.

### 2) EX-SITU TRAINING

In this section, we discuss the steps of ex-situ training algorithm. The goal of ex-situ training is to achieve high testing accuracy in hardware although a precursor DNN is trained in software. For this training, we also adopt weight quantization and allocate a separate memory in software where we store the high precision weights (similar to in-situ training) in addition to the DNN weights. The training algorithm shown in Fig. 2b remains the same for ex-situ training. After each iteration (forward and backward pass), high precision weights are updated and then quantized. Ideally, these quantized weights should be used as DNN weights for the next iteration in case of deterministic quantized neural network learning [1]. However, as we are dealing with a stochastic device for our inference engine, we include stochastic behavior of synaptic weights during learning. This stochasticity is obtained from a statistical distribution of the device (shown in Fig. 1d) rather than from uniform random distribution [1] or Gaussian distribution [38, 52]. For example, in 5-level quantization if the quantized weight is 0 then the DNN weight can be of any values selected randomly from the $< m_z >$ distribution of $K_u = 7.5 \ (\times 10^5) \ \text{J/m}^3$ which is responsible for generating quantized weight of 0 (see Fig. 1d).

The noise tolerance margin α is also used during ex-situ training. This will relax the stringent requirement of programming a stochastic DW to a predetermined learned weight value and potentially save a large number of programming attempts. More importantly, if the DNN becomes aware of the statistical distribution of the device during training it can perform well during inference as the same device based DNN is used for inference.

Once the ex-situ training is accomplished, the DNN weights (or the high precision weights) are quantized and transferred to the DW devices by suitable programming. Here, the learned weights and the programmed weights may not be the same due to the programming noise. During the programming, we allow the same noise tolerance margin, α that is used during training. Thus, any programmed device weight, $W_{ij}$ need to satisfy, $W_q - \alpha \leq W_{ij} \leq W_q + \alpha$ for a target-quantized weight of $W_q$. The devices can be programmed by repeated programming or performing read-verify-write operation in a loop, which is called "Open loop off device" method [53]. As we have already trained our network with stochastic distribution of weights by introducing finite α, the network is expected to perform well during testing when we allow the same noise tolerance level for programming the device.

### C. TESTING THE DNN
During the testing stage, we computed the predicted class for all the image samples from the MNIST test dataset using the trained DNN and compared it to the desired class. The percentage accuracy is calculated it by dividing the total number of accurate predictions to the total number of test samples. During the testing stage, we consider two scenarios depending on in-situ or ex-situ training. When both the training and testing is performed on simulated hardware, the testing accuracy we record is termed online testing accuracy. In contrast, when the training is performed off-line (ex-situ) in software and we program the hardware (simulated device in this case) prior to testing according to the learned weights then the testing accuracy we record is termed offline testing accuracy.

## IV. RESULTS AND DISCUSSIONS

### A. DNN CONFIGURATION SELECTION
The focus of our paper was to demonstrate the ability to classify images using a DW device based DNN and benchmark its performance against a DNN with floating precision (32-bit) weights. The topography of the benchmark DNN can be arbitrary as the inference accuracy vary widely across the spectrum of the parameters such as hidden layer number, layer size ratio (ratio of neurons between a layer and the next layer), learning rate constant (see Fig. S1 in supplementary). Thus, one can select multiple configurations for the DNN and achieve good accuracy. We select a benchmark DNN architecture consisted of a network with three hidden layers, an initial learning rate of 0.007 and a layer size ratio of $\frac{1}{2}$. We assume a learning rate decay of 10 % after each epoch and use stochastic gradient decent method as the optimizer. The selection criteria are detailed in the supplementary section. After training the selected benchmark DNN for 10 epochs, the test accuracy we achieve is 97.1 %. We note that there are opportunities to improve the accuracy further in terms of topography, batch normalization, dropout and selection of different optimizers. However, the main goal of this study is to show how well a stochastic and low precision DW based DNN can perform in comparison to a similar architecture floating precision DNN. The selected topography mentioned above is used throughout the study to implement the DW device based DNN.

### B. ONLINE (IN-SITU) TRAINING
After determining the DNN topography we investigate the test accuracies of the DNNs that are built from 2-state, 3-state and 5-state DW devices and trained them with the proposed online training algorithm. For simplicity, we did not consider additional hardware non-idealities that could arise from peripheral circuits or unresponsive devices as these factors would automatically be included as constraints during the on-line training [38] and would not result in a significant degradation in performance compared to our current work. The online training accuracy and online testing accuracy are plotted in Fig. 3 with the number of epochs for two different noise tolerance margins, $\alpha = 0.15$ (15% of maximum possible absolute weight) and $\alpha = 0.25$ (25% of maximum possible absolute weight) used during the training stage.

The effectiveness of the proposed in-situ training algorithm is evident from Fig. 3a and Fig. 3b which plots the in-situ training accuracies for different state devices for low ($\alpha = 0.15$) and high ($\alpha = 0.25$) noise tolerance margin respectively. The results are also compared with baseline accuracy (accuracy of a same topography DNN with floating precision weights and no stochasticity). The training accuracies for DW device based DNNs increase with the number of device states and almost reach the baseline accuracy of ~ 99.6 % for low noise tolerance of α=0.15 as can be seen from Fig. 3a. However, the training accuracies for DNNs with high noise tolerance, α=0.25 become slightly lower (see fig. 3b) as these networks allow higher deviation from the target quantized weights. Nonetheless, competitive training accuracies are achieved for both 3- and 5- state devices with high noise tolerance margin.

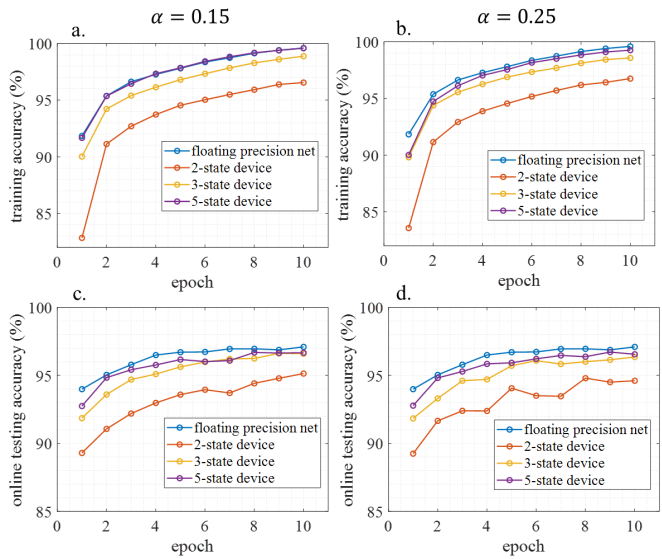

**FIGURE 3.** Online training accuracy and online testing accuracy for DNNs with different state DW devices for two different noise tolerance margins. These accuracies are compared with a DNN trained and tested with full precision weights and no stochasticity (baseline accuracy) a. and b. show the online training accuracies with the numbers of epochs for $\alpha$=0.15 and 0.25 respectively. c. and d. show online testing accuracies with numbers of epochs for $\alpha$ = 0.15 and 0.25 respectively.

After each epoch of the in-situ training we test the DNN with test images form MNIST dataset and compute the online test accuracy. Fig 3c and 3d plots online testing accuracies for low and high level of noise tolerance margin respectively. The baseline (DNN with floating precision weights and no stochasticity) test accuracies are plotted for comparison. For low noise tolerance margin of $\alpha$=0.15, the test accuracy is highest for 5-state device and reaches $\sim$ 96.67% after 10 epochs of training. This accuracy is very close to the baseline test accuracy of $\sim$ 97.1 %. It is important to note that, the 3-state device based DNN achieves a test accuracy of $\sim$ 96.6 % after 10 epochs of training, which is similar to a 5-state device. When the noise tolerance margin is increased to $\alpha$=0.25, the test accuracies for 5-state and 3-state devices are $\sim$ 96.56% and $\sim$ 96.36% after 10 epochs of training. Thus, a maximum decrease of accuracy of $\sim$ 0.74 % from 32-bit precision weight is recorded for a 3-state stochastic weight. We note that, the test accuracies for 2-state device are $\sim$ 95.14% and $\sim$94.64% for low and high noise tolerance margin respectively. Thus, the same topography networks for 2-state does not achieve comparable test accuracies. Changing the topography such as increasing the number of neurons in hidden layers can increase the accuracy of binary DNN [54].

Next, we analyze the total number of programming pulses that are applied to the DW devices during the course of the online training at various epochs. Because the network updates the high-precision weights, a single weight may have its high precision value updated many times before crossing the threshold to update the DW device weight. Because the number of device updates is dependent on the number of times a high precision weight crosses the threshold; the larger the threshold the fewer the updates. Between the 2, 3 and 5 state networks the 5-state has the smallest threshold, which increases the number of DW device updates as seen in Fig. 4. These DNNs are also compared with a DNN trained with floating precision weights (and no stochasticity). In floating precision DNN, all the weights are updated at each time a training image is passed to the network. Thus, although the network is better trained with increasing number of epochs, the weight update count remains almost constant as seen in Fig. 4. In contrast, for DNNs with limited state DW devices with the proposed training method, the programming instances decrease significantly with the number of epochs. As expected, with low noise tolerance margin the DNNs with DW devices become more selective and require higher number of weight updates during the course of the training (though this is much smaller than the case of floating precision weights).

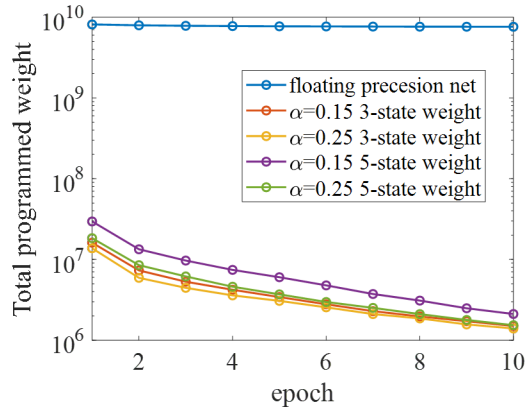

**FIGURE 4.** Comparison of the total number of programmed weights with the number of training epochs for different networks. A significantly lower number of weights are updated during the proposed online training compared to the full precision weight network of the same architecture.

In Fig. 5, we show the convergence of DW device weights during the training. DNN weights whose noise tolerance are higher will converge to a value quicker, on an average, than a weight with a lower noise tolerance. In Fig. 5a and 5b, the DW device weights fall within $\pm\alpha$ of the quantized weight value. In both cases, the DW device weight is closer to the high precision value than the quantized weight, which tends to provide a higher accuracy for our DW based DNN.

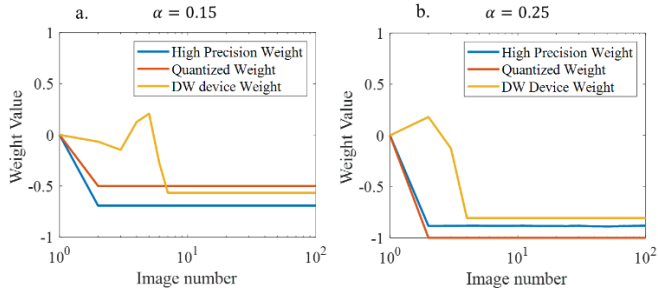

**FIGURE 5. Weight evolution of high precision weight, quantized weight and the DW device weight during the first few training images for two different noise tolerance margin a. $\alpha = 0.15$ 0.15 b. $\alpha = 0.25$. The synaptic weight shown here is connected between the two neurons located in hidden layer 2 and 3.**

## C. OFFLINE (EX-SITU) TRAINING

In this section, we first analyze the effectiveness of our proposed ex-situ training by comparing it with other techniques. For that, we train several precursor DNNs in software using different offline training algorithms (Fig. 6) and then test the DNNs, which are built from DW synaptic devices (3- state and 5-state hardware). Each of the DNNs are trained offline with a total of 10 epochs (train with entire training dataset 10 times) and prior to the testing the DW devices are programmed according to the weights that are learned offline. These results are shown in Fig. 6a and 6b when we consider a low ($\alpha = 0.15$) and high ($\alpha = 0.25$) value of noise tolerance margin to program the devices. In both Fig. 6a and 6b, for hardware test accuracies, a corresponding software accuracy is presented side by side with green bar. When the exact learned weights (no programming noise is considered while transferring the learned weights to the device) are used to test the DNNs we call it software accuracy.

When offline training is performed with both the floating precision and quantized weights cases, the testing accuracy is low, as can be seen from Fig. 6a. After floating precision weight training, the learned weights need to be converted to 3- or 5-state weights to program the DW devices. Thus, for both of the 3- and 5-state hardware the test accuracies degrade compared to software accuracy of $\sim 97.1\%$. Converting floating precision learned weights to 5-state compatible weights (5-level quantization) generates smaller deviations compared to the 3-state weight (3-level quantization). Thus the 5-state device provides higher test accuracy which is $\sim 87\%$ compared to the 3-state which is only $\sim 10\%$.

Training with quantized weights improves the test accuracies to $\sim 90\%$ for 5-state device (see Fig. 6a) as the network becomes aware about the limited states of the weights during the training period. However, the test accuracy remains low (software accuracy is $\sim 96.74\%$). The accuracy loss is mainly due to the deviation of the programmed weights from the learned weights. We note that, with floating precision training, weight deviations occur in two ways: converting the floating precision weights to quantized weights and during the programming of the device where the target quantized weights are not achieved deterministically. However, with quantized training only the latter deviation occurs during the testing stage.

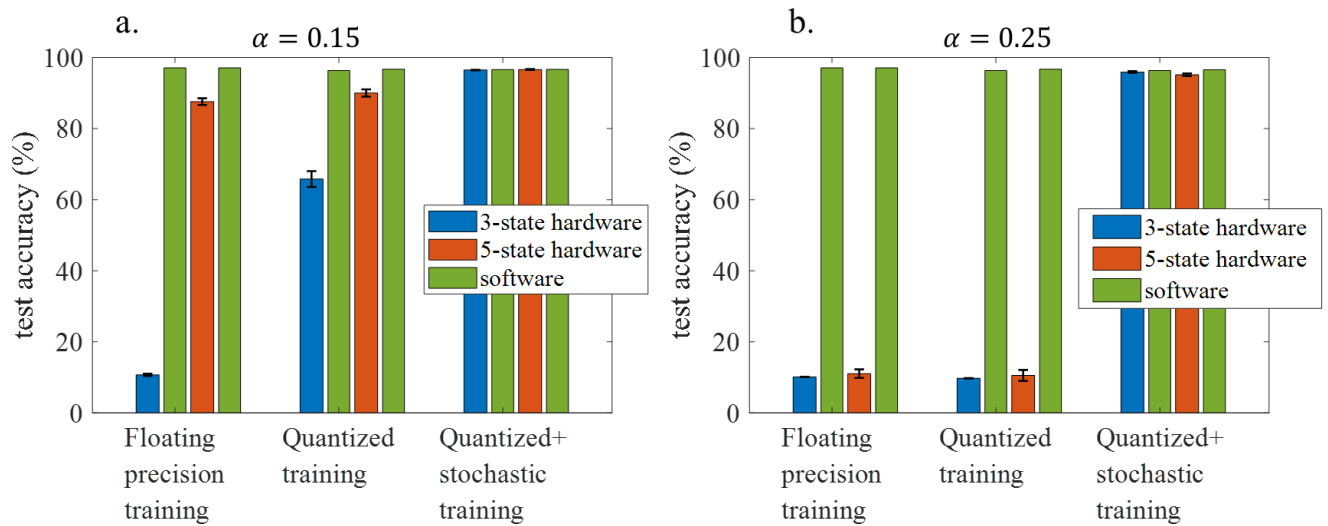

**FIGURE 6. Testing accuracy comparison of 3-state and 5-state DW device based DNNs for different ex-situ training algorithms for a programming noise toleration window of a. $\alpha = 0.15$ b. $\alpha = 0.25$. The networks are trained off-line with floating precision weights, quantized weights, and stochastic quantized weights derived from micromagnetic simulation. Each of the network is trained with a total 10 epochs. Once training is done, the 3-state and 5-state DW devices are programmed based on the quantized value of trained weights prior to testing. For different training algorithms and for each of the test accuracy of DNN built from 3- and 5-state hardware, a corresponding software test accuracy (no programming noise is considered and exact trained weights are used for testing the DNN) is plotted side by side with green bar. Error bar seen in the figure is calculated for a total of 10 test trials. For both noise tolerance windows, the test accuracy is highest when trained with proposed training algorithm (quantized + stochastic).**

In contrast, with our proposed training which we call quantized + stochastic training, the test accuracy increases and reaches up to $\sim$ 96.63% for 5-state device, which is close to the software accuracy of $\sim$96.67%. The accuracy improvement can be attributed to even smaller deviation of the programmed weights from the learned weights. Unlike quantized training, in our proposed training the weight quantization is also accompanied by training the DNN weights according to the statistical distribution of the device. As a result, during back propagation, the high precision weights are updated depending on the weighted sum performed over the imprecise DNN weights (which are mapped from the stochastic distribution of the device as in Fig. 1d). In other words, the high precision weights are being tuned based on the stochastic signature of the device. Thus, the statistical distribution of the device is embedded in the learning. When the same devices are used for testing, the distribution matches better and this plays an important role for improving the test accuracy. This finding is also supported by other works [38, 52]. Ref. [38] shows that the DNN trained with Gaussian distributed weights of a certain standard deviation performs better when weights of same standard deviation are used for inference.

With high programming noise, for both floating precision and quantized training, the programmed weights deviate more from the learned weights because of the higher noise tolerance. Thus, the test accuracies for 3- or 5-state hardwires degrade significantly compare to the software-based accuracies as seen from Fig. 6b. In contrast, with our proposed training method, the DNNs are made aware about the statistical distribution of the device thus resulting in significantly higher test accuracy compare than other offline training methods. (Note that as the device statistics are not Gaussian and instead heavily dominated by the pinning positions, training with Gaussian distributed weights does not improve accuracy and was not employed).

We also studied the evolution of offline test accuracies with the number of epochs for different state devices, which are presented in supplemental Fig. S3. The influence of noise tolerance margin $\alpha$ on training accuracy, online testing accuracy and offline testing accuracy for DNN with limited state device (5-state) is shown in supplementary Fig. S4, which shows that off-line testing accuracy is affected most by the choice of different $\alpha$.

### D. ENERGY DISSIPATION

Energy dissipation to program a DW-synapse depends on charging the piezoelectric layer by applying a voltage pulse, $\frac{1}{2}CV^2$ as well as $I^2R$ loss due to the SOT current in the heavy metal layer. The maximum change in PMA is $\Delta PMA = 0.5 \times 10^5$ $J/m^3$. For magnetic racetrack of CoFe the saturation magnetization is, $\lambda_s$=250 ppm. Using the above values, the required maximum stress, $\sigma$ is calculated to be, $\frac{\Delta PMA}{\frac{3}{2}\lambda_s}$=133 MPa. For CoFe with Young's Modulus of 200 GPa, the required strain is, $\frac{133\ MPa}{200\ GPa}$ $\sim 10^{-3}$. Previous study

[46] showed that $10^{-3}$ strain is possible in Lead Zirconate Titanate (PZT) piezoelectric with an applied electric field of E=3 MV/m when the electrode dimensions are in the same order of the PZT thickness. If we consider PZT layer to be b=60 nm thick (same as top electrode or racetrack width as shown in Fig. 1(b)) then a voltage of, E*b = 0.18 V applied between the top electrode pair and the bottom electrode can generate the required strain. If the top electrode length L=600 nm (same as racetrack length 600 nm) and width b=60 nm is considered, and relative permittivity of PZT is $\epsilon_r$=3000 then the effective capacitance is calculated to be $\frac{\epsilon_0\epsilon_r(L*b)}{b} \sim 16$ fF. This suggests a $\frac{1}{2}CV^2$ loss of ~0.5 fJ considering two top electrodes on both sides of the racetrack.

The heavy metal layer is considered to be Pt and for $600\times60\times$ $nm^3$ dimension Pt layer the resistance is calculated to be 200 $\Omega$ assuming the resistivity of Pt to be 100 $\Omega$nm. The heat loss in the heavy metal layer is calculated to be 2.2 fJ for a fixed SOT generating current pulse of magnitude $35 \times 10^{10}\ A/m^2$ applied for 1 ns. Thus, the total energy dissipation to program a synapse is calculated to be 2.7 fJ.

#### 1) IN-SITU TRAINING
With in-situ training, highest inference accuracy is achieved for a 5-state device when a low noise margin is considered during the training. However, with higher noise tolerance margin similar test accuracy is obtained with fewer device updates as can be seen from Fig. 4. For 5-state device, if we consider a noise tolerance margin of $\alpha$=0.25, the total number of weight updates are calculated to be $\sim$ 48 million after running the training for 10 epochs. *Thus, the energy dissipation to program the DNN synapses is calculated to be ~13pJ for one inference event followed by the weight updates* (10000 test images in MNIST).

#### 2) EX-SITU TRAINING
With ex-situ training, highest inference accuracy is achieved for 5-state device when the noise margin to program the DW devices is considered to be low. Fig. 7 shows the cumulative probability of the DW device weights for different programming condition for a 5-state device. The solid black line represents the target quantized weights of 1, 0.5, 0, -0.5 and -1 (in this case -0.833) which can be achieved by a combination of fixed SOT current pulse and a varying amplitude voltage pulse which modulates the anisotropy of the racetrack to $K_u$ = 7, 7.25, 7.5, 7.75 and 8.0 ($\times 10^5$) J/m3 respectively. The adjacent red dotted lines in the figure shows the noise margin ($\alpha = 0.15$) that is allowed while programming the DW device to a specific quantized state. From Fig. 7 it can be seen that the probability of programming the DW device weight to a quantized value of 1 is the lowest which is $\sim$ 6 % meaning a number of $\sim$ 20 attempt is required to program the device. If we consider the worst-case scenario, then after ex-situ training prior to the inference we need 20 programing pulses to program each of the DW devices implementing the DNN weights. *Thus, for our network topology of 784-392-196-98-10 neurons, the energy* *dissipation to program the DW synapse is 2.8 pJ per inference event.*

The energy dissipation to program the DW devices in in-situ training is found to be 5× the dissipation incurred in ex-situ training, which is a moderately low provided that the training is performed over the entire 60000 training images for 10 times. This low dissipation in-situ training is possible due to distinct features of proposed training algorithm that benefits from weight quantization and noise tolerance margin. Large interstate interval in quantized learning helps to reduce the number of weight updates. Moreover, once the device is programmed within the noise tolerance margin, further write operation is avoided with a simple low cost read operation. We note that onsite learning is attractive in power constraint edge devices, where the learning itself needs to adapt and respond to a continuously evolving environment. Embedded medical systems [55], real time intrusion detection [56], and dialect specific speech recognition systems can be benefitted from such onsite learning. Ex-site learning can perform inference tasks in edge devices with energy efficient manner (given the training is performed over cloud server), however the benefit can only apply to non-adaptive tasks.

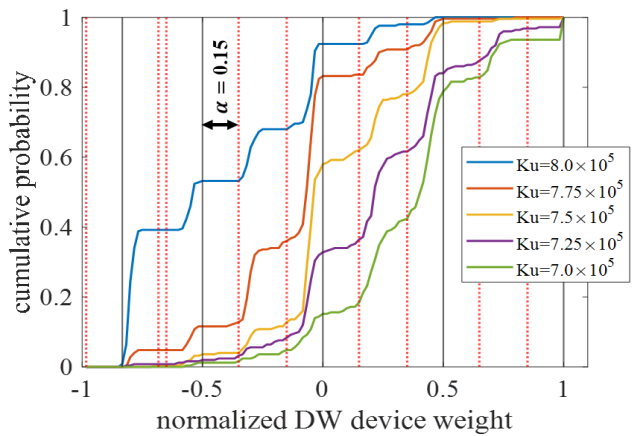

Finally, when paired with purely CMOS based device, spintronic devices can provide the non-volatility with higher area efficiency and less static power dissipation. Non-volatile PCM and RRAM technology can provide smaller footprint device, however the energy dissipation to program such a synaptic device is in the ~pJ range [23, 54]. Moreover, these devices have low endurance compare to spintronic devices [2].

## V. CONCLUSION
We have shown that DNNs with extremely low resolution and stochastic DW device-based synapses can achieve high classification accuracy when trained with appropriate learning algorithms. In this study, both in-situ and ex-situ training algorithms are presented for DNNs that are implemented with 2-state, 3-state and 5-state DW devices. For in-situ training, a high precision memory unit is employed to preserve and accumulate the weight gradients, which are quantized to obtain target conductance for updating the low precision DW devices. A noise tolerance margin further allows for random deviations of the programmed conductances from the target conductance values. For ex-situ training, a precursor DNN is first trained in software by performing weight quantization and considering a noise tolerance margin from the quantized weight and later tested with an equivalent DNN of DW devices programmed with the same noise margin. While the energy dissipation statistics for programming the DNN synapses shows that ex-situ method is energy efficient, the in-situ training provides an opportunity to learn and adapt to changing environment with only 5× more dissipation (despite the fact that the in-situ training is performed over a vast number of training images for many epochs). This technology is specifically attractive for low power intelligent edge devices of future IoT where energy requirement is at a premium.

## ACKNOWLEDGMENT

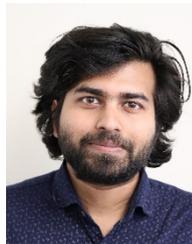

**WALID AL MISBA** received the B.Sc. degree in electrical and electronic engineering from the Bangladesh University of Engineering and Technology, Dhaka, Bangladesh in 2013. He received the MS in electrical engineering from Tuskegee University, Alabama, US. He is currently pursuing his Ph.D. degree in Mechanical and Nuclear Engineering with Virginia Commonwealth University, Richmond, VA, USA.

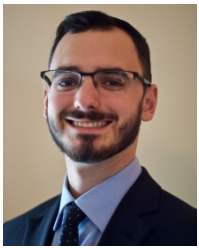

**MARK LOZANO** received the B.S. degree in mechanical engineering from the Virginia Commonwealth University, Richmond VA, in 2020. His research interest includes designing machine learning algorithms for varying applications. After fulfilling his ongoing contract in Chantilly VA, he plans on going back to school to further research in neuromorphic computing and obtain a M.S. in computer science.

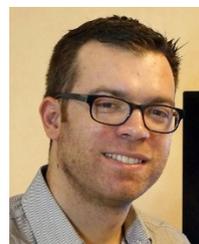

**DAMIEN QUERLIOZ** (M'08) received the predoctoral education from the Ecole Normale Supérieure, Paris, and the Ph.D. degree from Université Paris-Sud, in 2008. After postdoctoral appointments at Stanford University and CEA, he became a Permanent Researcher with the Centre for Nanoscience and Nanotechnology, Université Paris-Sud. He focuses on novel usages of emerging non-volatile memory, in particular relying on inspirations from biology and machine learning. He coordinates the INTEGNANO Interdisciplinary Research Group. He is currently a CNRS Research Scientist with Univeristé Paris-Sud. In 2016, he was a recipient of the European Research Council Starting Grant to develop the concept of natively intelligent memory. In 2017, he received the CNRS Bronze Medal.

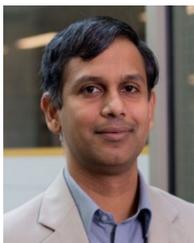

**JAYASIMHA ATULASIMHA** Jayasimha Atulasimha (SM'11) received the M.S. and Ph.D. degrees in aerospace engineering from the University of Maryland, College Park, MD, USA, in 2003 and 2006, respectively. He is a Professor of Mechanical and Nuclear Engineering and Electrical and Computer Engineering with Virginia Commonwealth University, Richmond, VA, USA. His current research interests include magnetostrictive materials, nanoscale magnetization dynamics, and multiferroic nanomagnet-based computing architectures.

Supplemental Material to

# Energy efficient learning with low resolution stochastic Domain Wall synapse based Deep Neural Networks

Walid A. Misba[1], Mark Lozano[1], Damien Querlioz[2] and Jayasimha Atulasimha[1,3]

[1]Mechanical and Nuclear Engineering Department, Virginia Commonwealth University, Richmond, VA 23284 USA

[2] Université Paris-Saclay, CNRS, Centre de Nanosciences et de Nanotechnologies, Palaiseau, France

[3]Electrical and Computer Engineering Department, Virginia Commonwealth University, Richmond, VA 23284 USA

## S1. Selection of Deep Neural Network (DNN) Architecture:

The network topography we try to optimize in terms of training and testing accuracy used floating precision (32-bit) weights for the synapses. This optimized network was used to benchmark the performance of the DW based DNN of same topography.

To select the optimal network topography multiple characteristics are considered: training accuracy, testing accuracy, network size, and fitness. An algorithm was designed to create, train, and test networks with each having a unique combination of parameters such as hidden layer numbers ranging from 0 to 9, the layer size ratio (ratio of a layer's neurons to the previous layer neurons) ranging from 1/32 to 32/32, and a learning rate ranging from 0.001 to 0.009. Parameters that remain consistent are the learning rate decay and the starting epoch of learning rate decay. All the networks are trained using only one epoch, which results in underfitting networks where the training accuracy was lower than the testing accuracy. Limiting the number of epochs to one saves computational burden and as the network topographies are compared to one another so rather than achieving independent high accuracies the accuracy in reference to an alternate topography is prioritized. The two parameters learning rate decay and the starting epoch of decay, are chosen to be 10 % at each epoch and the first epoch. The networks are trained on the full set of 60,000 training images and tested on the full set of 10,000 testing images from the MNIST handwritten digits database. From hidden layer number variation results (not shown here) we find that the accuracy increases with the increase of hidden layer, however, when the hidden layer number is greater than 3 then the accuracy does not increase appreciably. So, we select a total of 3 hidden layers for our DNN configuration. Fig. S1 shows the height map for training accuracy, testing accuracy and average of training and testing accuracies for different architectures of DNN with 3-hidden layers. We can see from Fig. S1 that the accuracies vary widely across the spectrum of variables. Thus, one can select several configurations of DNN and achieve good accuracies. We choose a layer size ratio of 16/32, so for the selected network the number of neurons for the first hidden layer is $\frac{1}{2} \times$ the number of neurons of the input layer and so on. We choose a learning rate constant of 0.007 as it would provide sufficient learning capacity with the increase of training epoch (learning rate decays by 10 % at each epoch). Thus, the final architecture consists of a network with three hidden layers, an initial learning rate of 0.007 and a layer size ratio of $\frac{1}{2}$.

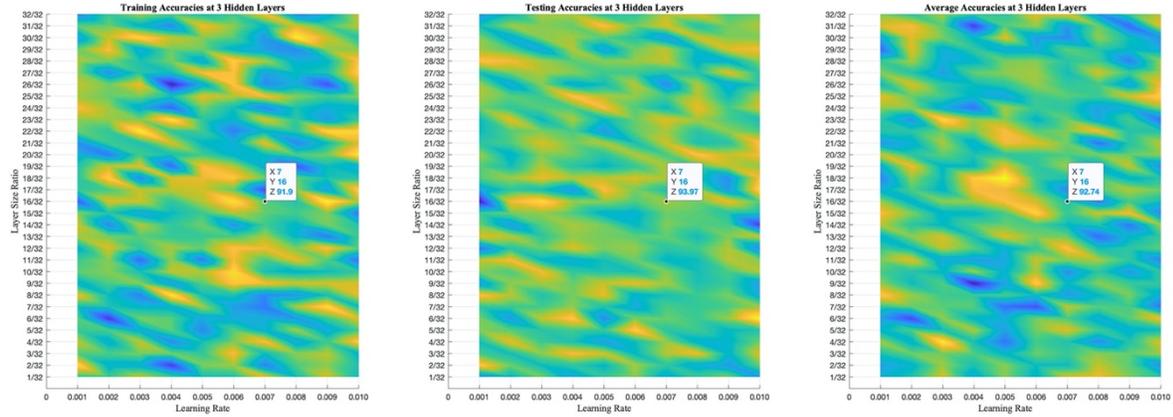

Fig. S1: Height maps showing the accuracies of networks with varying topographies with 3 hidden layers. The highlighted data point is the final topography chosen for our DW device based DNN as it is small in number of synapses. From left to right are the networks training accuracies, testing accuracies, and the average between the two.

## S2. Equilibrium DW Position Distribution with respect to Dominant Pinning Sites (Notches):

The following Fig.S2b.-f. shows the equilibrium DW position distribution for the racetrack shown in Fig. S2a for five different programming conditions represented by $K_u$ = 8, 7.75, 7.5, 7.25 and 7.0 ($\times 10^5$) J/m3. Although DWs positions are pinned stochastically due to the edge roughness ($\sim$ 2nm rms) and the thermal noise due to temperature, T=300 K, however, most the DWs are pinned at the location of the pining sites. Thus, the distribution is heavily dominated by the pining site locations.

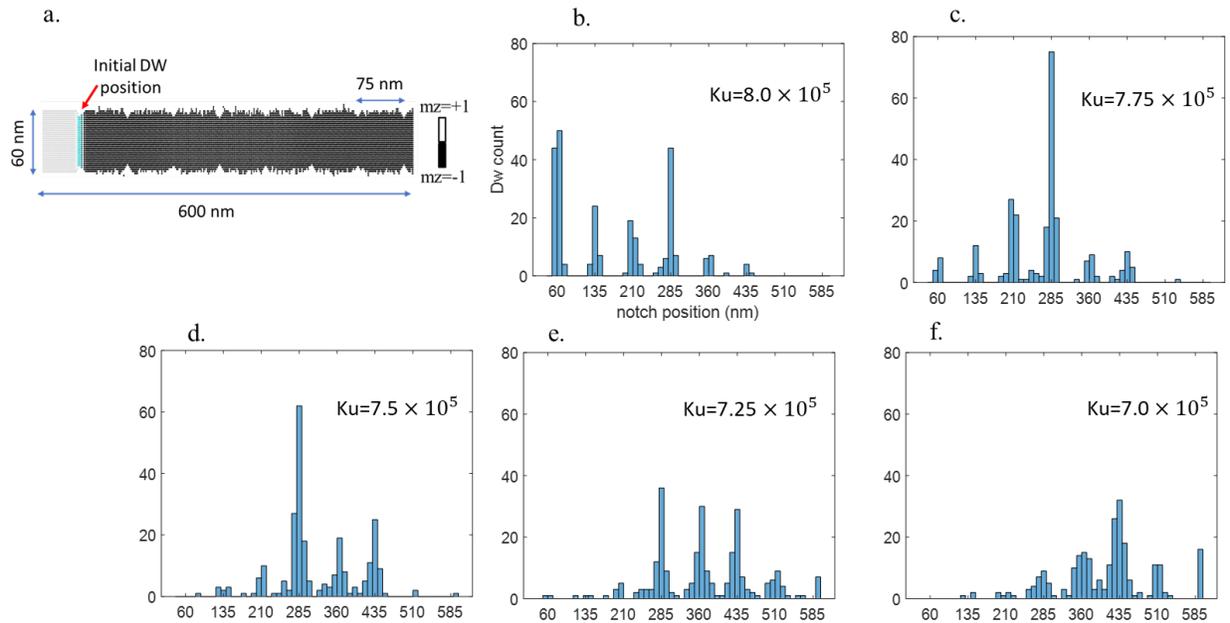

Fig. S2: a. Racetrack of dimension 600 nm x 60 nm x 1nm with rms edge roughness of $\sim$ 2nm and the corresponding DW at the initial position that is 60 nm from the left end. Engineered notches starting from 60 nm are placed at a regular interval of 75 nm in the racetrack. b.-f. Distribution of equilibrium DW positions along the racetrack of Fig. S2a for different programming conditions

represented by different PMA coefficient $K_u$. The DWs are primarily pinned at or around the notches, thus the distribution become skewed to different notch positions for different programming conditions.

## S3. Evolution of off-line Testing Accuracies with training Epoch:

The off-line testing accuracies with the number of epochs are presented in Fig. S3a and Fig. S3b for both low and high noise tolerance levels. The baseline testing accuracies with floating precision weights are also plotted for comparison. After each epoch of the training, the trained weights are collected and the devices in the crossbar are programmed. As we have allowed noise tolerance margin during testing (same noise tolerance margin $\alpha$ that is used during training) to program the device so for different trials of testing, the programmed weights of the network could be different. So, we consider a total 10 different trials for testing the DNNs. The error bar shown in the figure is computed for 10 trials of testing after each epoch of the training. For noise tolerance margin of $\alpha$=0.15, the offline testing accuracies reaches close to the baseline testing accuracies (same topology DNN with floating precision weights and no stochasticity) after 10 epochs of training for 5-state and 3-state devices as seen from Fig. S3a. The highest offline test accuracy of ~ 96.63% is achieved for 5-state device which is very close to the baseline test accuracy of ~ 97.1%. However, for $\alpha$=0.25, the test accuracies degrade for all the devices as can be seen from Fig. S3b. The highest offline test accuracy for 5-state, 3-state and 2-state devices are obtained after 10 epochs of training which are ~ 95.14%, ~95.93% and ~ 94.24% respectively.

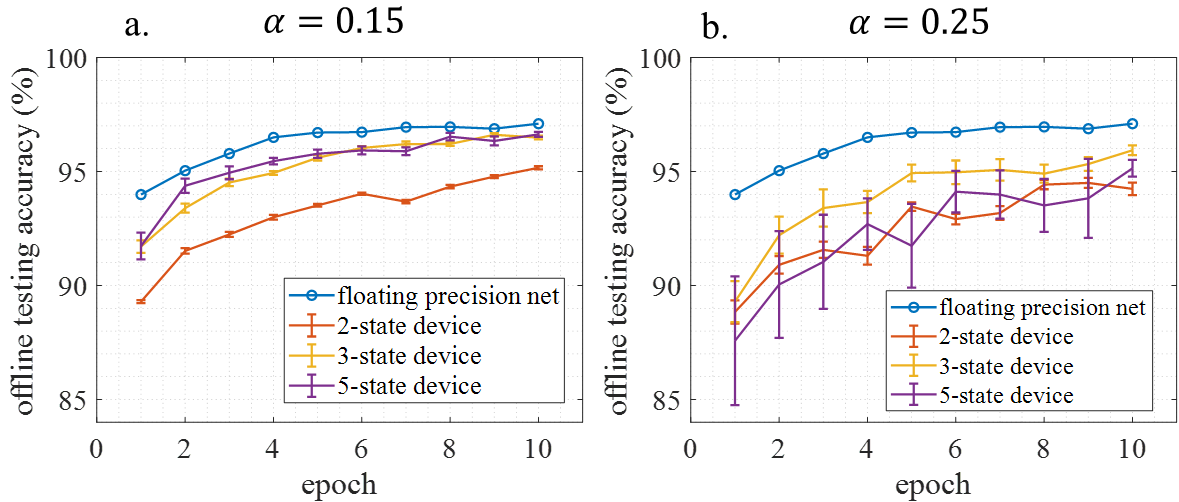

Fig. S3. Offline testing accuracy for DNNs of different state DW devices for two different noise tolerance margins, $\alpha$. The accuracies are compared with a DNN trained and tested with full precision weights and no stochasticity (baseline accuracy). Offline testing accuracies with the numbers of epochs for noise tolerance margin of **a.** $\alpha$ = 0.15 and **b.** $\alpha$ = 0.25 used during the programming of the devices. Error bar is calculated for a total of 10 trials at each of the data points.

## S4. Influence of Noise Tolerance margin on Accuracy:

To examine the influence of noise tolerance margin on accuracy, training accuracy, online testing accuracy (training is done in-situ) and offline testing accuracy (training is done ex-situ) of 5-state DW based DNNs are compared in Fig. S4 for high and low level of noise tolerance margin, $\alpha$. From Fig. S4a and S4b we can see that, the difference in noise tolerance does not appear to have a significant effect on the training or online testing accuracies. This is due to the fact that backpropagation is performed over the imprecise DW device weight which is then used to update the high precision weights for the network. In effect the potential values selected at random for the DW device weights are then known by the network, and through training change the DW device weights accordingly based on the quantized value of the high precision weights. Once trained in-situ, the learned weights remain the same during testing due to non-volatility. In contrast, after ex-situ training the learned weights are transferred to the DNN by programming the DW device with a noise tolerance margin that is used during the training. When the devices are programmed prior to testing, programming noises are added to the DW device weights. The higher the noise tolerance margin, the higher the amount of noise that could be added to the programmed device weights. Thus, the offline testing accuracy degrades with higher noise tolerance margin. For example, if the trained weight is 0.24 then the quantized weight would be 0 and after programming the DW device weight could be -0.24. Thus, with noise tolerance margin of $\alpha=0.25$, the maximum deviation of the programmed weight from the learned weights could be $2\alpha=0.5$ whereas for low noise tolerance margin $\alpha=0.15$ the maximum deviation could be 0.3.

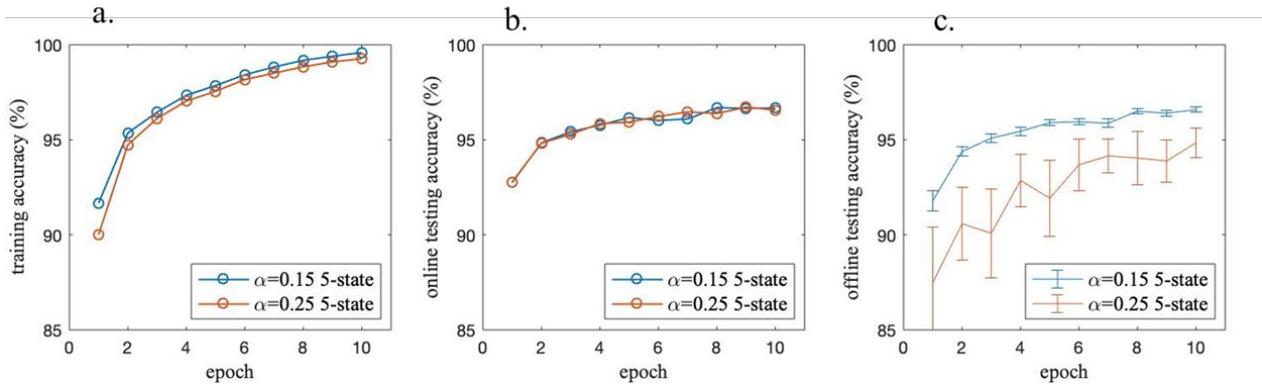

Fig. S4. a. Training accuracy b. Online testing accuracy and c. Offline testing accuracy for a 5-state DW device based DNN for two different noise tolerance margins of $\alpha$. The training accuracy and online testing accuracy does not change appreciably for different noise tolerance margin. Ex-situ testing accuracy decreases with high noise tolerance margin due to the higher deviation of device weights during the programming of the devices.